# Is the Universe Uniquely Determined by Invariance Under Quantisation?

Philip E. Gibbs[1]

## Abstract

In this sequel to my previous paper, "Is String Theory in Knots?" I explore ways of constructing symmetries through an algebraic stepping process using knotted graphs. The hope is that this may lead to an algebraic formulation of string theory. In the conclusion I speculate that the stepping process is a form of quantisation for which the most general form must be sought. By applying the quantisation step a sufficient number (possibly infinite) of times we may construct an algebra which is equivalent to its own quantisation.

[1] e-mail to philip.gibbs@eurocontrol.fr

## Introduction

In the previous e-print (Gibbs 1995) I defined a Lie superalgebra on a basis of discrete loops which I interpret as the Fock space for strings made of fermionic partons strung together. This symmetry was derived on the basis of the principle of event-symmetric space-time (Gibbs 1996a, 1996b). I speculated that the construction might be generalised to give an algebraic theory of superstrings which theorists expect to exist (e.g. Schwarz 1995).

The relations used in definition of the discrete string theory were noticed to be similar to the form of Skein relations in knot theory. This suggests a link between knot theory and string theory which I will pursue further in this paper. The construction remains incomplete but has led to some interesting possible conclusions about the role of quantisation which are worth discussing.

## Discrete Superstring Symmetry

First of all I will recall the definition of the discrete string symmetry

Let $E$ be a set of $N$ space-time events and let $V = span(E)$ be the $N$ dimensional vector space spanned by those events. Then define $T = Tensor(V)$ to be the free associative algebra with unit generated over $V$. The components of $T$ form an infinite family of tensors over $V$ with one representative of each rank.

$$\Phi = \{\varphi, \varphi_a, \varphi_{ab}, \varphi_{abc}, \ldots\}$$
$$\Phi^1 \Phi^2 = \{\varphi^1 \varphi^2, \varphi^1 \varphi^2_a + \varphi^1_a \varphi^2, \varphi^1 \varphi^2_{ab} + \varphi^1_a \varphi^2_b + \varphi^1_{ab} \varphi^2, \ldots\}$$

The basis of this algebra already has a weak geometric interpretation as open strings passing through a sequence of events with arbitrary finite length. Multiplication of these strings consists merely of joining the end of the first to the start of the second. We can denote this as follows,

$$\Phi = \varphi + \sum_a \varphi_a a + \sum_{a,b} \varphi_{ab} ab + \sum_{a,b,c} \varphi_{abc} abc + \ldots$$

We now construct a new algebra by adding an extra connectivity structure to each string consisting of arrows joining events. There must be exactly one arrow going into each string and one leading out. This structure defines a permutation of the string events so there are exactly $K!$ ways of adding such a structure to a string of length $K$.

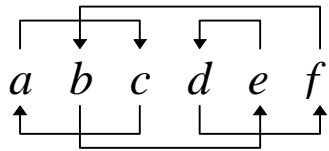

These objects now form the basis of a new algebra with associative multiplication consisting of joining the strings together as before, while preserving the connections.

Finally the algebra is reduced modulo commutation relations between events in strings which are defined schematically as follows,

$$a\ b\ +\ b\ a\ =\ 2\delta_{ab}$$

These are partial relations which can be embedded into complete relations. Closed loops which include no events are identified with unity. For example, the lines can be joined to give,

$$a\ b\ +\ b\ a\ =\ 2\delta_{ab}$$

This example shows the cyclic relation on a loop of two events. The arrows can be joined differently to give another relation,

$$a\ b\ +\ b\ a\ =\ 2\delta_{ab}$$

which is the anti-commutation relation for loops of single events.

By applying these relations repeatedly it is possible to reorder the events in any string so that the strings are separated into products of ordered cycles. Therefore we can define a more convenient notation in which an ordered cycle is indicated as follows,

$$(ab\ldots c)\ =\ a \rightarrow b \rightarrow \ldots \rightarrow c$$

We can generate cyclic relations for loops of any length such as,

$$(ab) = -(ba) + 2\delta_{ab}$$
$$(abc) = (cab) + 2\delta_{bc}(a) - 2\delta_{ac}(b)$$
$$(abcd) = -(dabc) + 2\delta_{cd}(ab) - 2\delta_{bd}(a)(c) + 2\delta_{ad}(bc)$$

and graded commutation relations such as,

$$(a)(b) + (b)(a) = 2\delta_{ab}$$
$$(ab)(c) - (c)(ab) = 2\delta_{bc}(a) - 2\delta_{ac}(b)$$

## Supersymmetry Ladder

The next stage of the algebraic string theory program is to construct a ladder operation which takes us from one supersymmetry algebra to another one. Starting from the one dimensional string supersymmetry constructed in the previous section, the ladder operator will take us up to a symmetry of two dimensional would sheets. Further steps take us up to higher dimensional algebras.

For simplicity we start with an ordinary Lie algebra whose elements satisfy the Jacobi relation,

$$[[A,B],C]+[[B,C],A]+[[C,A],B]=0$$

A new algebra is constructed by stringing these elements in a sequence and attaching them with an orientated string passing through each one like before. A difference introduced this time is that the string is allowed to have trivalent branches. To keep everything tidy we will have the strings going up the page and returning through a cyclic boundary.

To make it more interesting we must factor out the following relations,

$$A \; B \quad - \quad B \; A \quad = \quad [A,B]$$

The crossing lines do not (yet) indicated that the lines are knotted. It does not matter which goes over the other.

When we check the result of combining the interchanges of three consecutive elements as we did before, using the commutation relation above, we find that the result is consistent with the Jacobi relation provided we also apply the following associativity relations and the corresponding co-associativity relation for strings which join.

This process defines a new algebra like before except that now we start from any Lie algebra and create a new associative algebra. A new Lie algebra is then defined using the commutator of the algebra as the Lie product.

The algebra is significantly simplified by applying extra graph relations to remove holes.

This construction generalises easily to the Lie superalgebra case using graded commutator and Jacobi relations. Thus we have a ladder operators which maps one superalgebra to a new one.

In the case where we start with the discrete string super algebra the loops can be visualised as circling the new network. These can then be interpreted as sections of a branching string world sheet. The new algebra is therefore a symmetry of string world sheets. Application of the ladder operator increases the dimension of the structures each time.

The original Lie algebra is isomorphic to a subalgebra of the higher dimensional one. That is the subalgebra formed by simply looping each element to itself. Furthermore, there is a homomorphism from the higher algebra to the universal enveloping algebra below which is defined by removing the string connections.

It is possible to apply the ladder operator any number of times. In fact, since the old algebra is contained in the new, it is also possible to define an algebra generated by an infinite number of applications of the ladder operator which then contains all the lower ones. This largest algebra has the property that applying the ladder operator generates a new one which is isomorphic to the original.

This raises an interesting question. Is it possible that after only a finite number of applications of the ladder operator, you already arrive at the most complete algebra? Further steps may just create algebras isomorphic to the previous one. This is a difficult open question of (possibly) considerable physical significance.

## Knotted Algebras

The above superstring symmetries are all very well except that strings are not made from discrete fermionic partons. They are defined as continuous loops, but at the same time they may be topological objects which can be determined by discrete points. To try to capture this algebraically it may be necessary to envisage a string as being made from discrete partons with fractional statistics like anyons or parafermions. Such partons may be repeatedly subdivided into partons with smaller fractional statistics until a continuous limit is found. If strings are truly topological, an infinite sequence of subdivisions may not be necessary.

Motivated by these thoughts it would be natural to seek some kind of deformation of the fermionic string algebra replacing the sign factors in the exchange relations with some general $q$-parameter. It is also natural to replace the loops which connect the partons with knots. In doing so we immediately hit upon a fortuitous coincidence. The construction of invariant knot polynomials makes use of relations which are similar in certain ways to those we have already used. e.g.

$$q \times - q^{-1} \times = z \,||\,$$

This is the relation which defines the HOMFLY polynomial, but how can the discrete string algebras be generalised to incorporate these Skein relations?

## Generalised Symmetry

To be more systematic it might be better to consider generalised concepts of symmetry which cover fractional statistics in the same way as super-symmetry works for fermions and bosons. It is now well known that symmetry can be generalised through the use of quantum groups but super-symmetry is not itself a special case of a quantum-group.

A simpler generalisation of symmetry might be taken as simply the combination of an associative unital algebra and an R-matrix or Yang-Baxter operator. The algebraic operations can be displayed graphically as follows

multiplication: 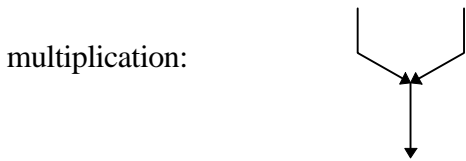

unit: 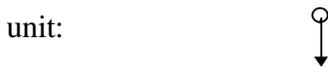

R-matrix : 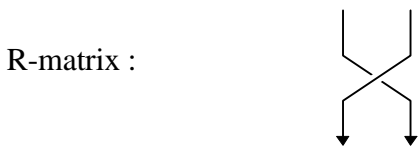

Algebraic relations such as associativity and the Yang-Baxter relations can then be shown graphically in the usual way. Another important relation showing consistency between multiplication and the R-matrix can be shown as follows

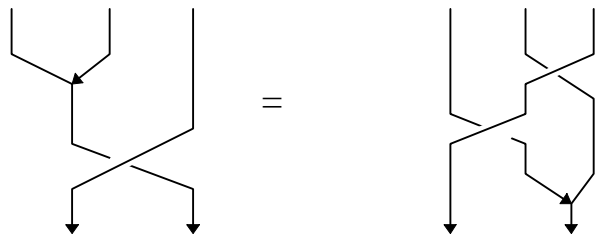

This combination defines an algebra which can be regarded as a generalisation of a Lie superalgebra, which would correspond to the special case where the R-matrix exchanges elements of the algebra while multiplying by sign factors corresponding to the parity of statistics. In the more general case the R-matrix does not square to unity and it is not appropriate to define a graded commutator.

Just as a $Z_2$ graded associative algebra can be used to construct a Lie superalgebra, a $Z$ or $Z_n$ graded algebra can be used to define a more general symmetry algebra in the sense above. It is merely necessary to define the R-matrix as commutation with a factor which is an $n^{th}$ root of unity. i.e.,

$$R: A \otimes B \to q^{grad(A)grad(B)} B \otimes A$$
$$q^n = 1$$

## Generalised Ladder

This generalised definition of symmetry is of interest because there is a way to construct a generalised ladder operation like the ladder of superalgebras. We must also stipulate that a scalar product can be defined through a symmetric bilinear operator. This can also be used to define a mapping between the algebra and its dual, to raise and lower indices. The structure constants of the multiplication must be cyclically symmetric with respect to this operator. There will then also be a co-unit, and co-product.

Graphically this means that the operations are duplicated with diagrams turned upside-down.

co-multiplication: 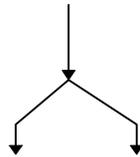

Using elements of an algebra of this type we define a new algebra based on strings of elements connected by trivalent graphs as before except that now when a string crosses another it is considered significant which goes over the other. In place of commutation relations we use the following

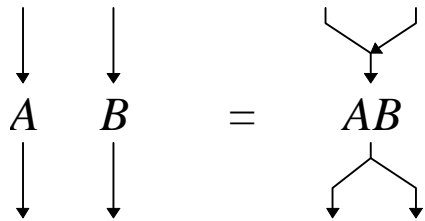

The associativity of the graph now corresponds directly to the associativity of the algebra. We can extend this to co-associativity which is indicated as follows using a Sweedler notation

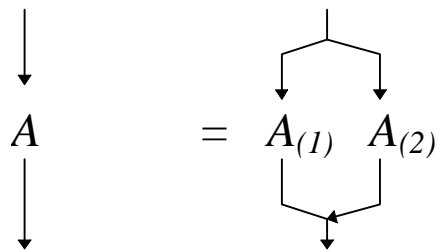

Finally we complete with a relation which uses the R-matrix

$$A \; B \;\; = \;\; R \;\; B \;\; A$$

In this relation the R-Matrix should be taken to be acting on both the algebra and its dual part in such a way as to be consistent with the direction of the twists. The switch of order of *A* and *B* is just symbolic of the R-matrix operation. Note that the strings get twisted together.

Together these relations define a new associative algebra with string diagrams placed side by side to form multiplication. Furthermore, it is possible to define an R-matrix operator on this algebra by applying the R-matrix of the previous algebra multiple times in order to move one element past another. Note that in this operation the R-matrix needs to be applied in a sense that does not twist the strings together.

This completes the construction of the ladder steps. It remains to define the first rung.

## Anyonic Loops

Is it possible to generalise the discrete string symmetry algebra to give an algebra of anyonic loops which would then serve as a first rung on the ladder?

This is not so easy to see. I have already speculated that the Skein relations of the HOMFLY polynomial may be a part of the solution and next I will show how this can come in.

Let us suppose that we have some R-matrix which acts on the tensor product of a vector space *V*.

$$R: \;\; V \otimes V \to V \otimes V$$

Now consider the algebra of endomorphisms on V which can also be regarded as being the same as the tensor product of the *V* with its dual

$$E = V \otimes V^*$$

There is a natural extension of the R-Matrix of *V* to an R-Matrix of $V^*$ and to $E$. This leads to a construction of the kind of symmetry algebra we are looking for with composition of the endomorphisms as multiplication.

If the R-Matrix is of Hecke type, then the relations of the first step of the ladder up from this one will be consistent with familiar Skein relations on the strings.

## Quantisation as Symmetry

The picture of a complete theory is becoming clearer. It will depend on an algebraic ladder which takes us step by step through the scale of increasing dimensions. If necessary the ladder can be climbed through an infinite number of steps leading to an algebra which is isomorphic to itself after applying the step one more time. We may find that, in fact, the algebra is complete already after only a finite number of steps.

If this stepping process is so important then we should try to understand what it is. one interpretation is in fact quite evident. The appearance of Feynman diagram structures and Fock spaces at each stage suggests that the step is itself a form a quantisation process, going from a classical system to a quantised version of it, then to a second quantised theory, a third quantised and so on. In that case the final system would have the property that it is dually related to its own quantisation, a property which is already thought to be a feature of string theory (Duff 1994).

To go further it will be useful to try to understand the most general processes of algebraic quantisation. Boulatov has used the quantisation of functions on a quantum group as a pre-theory for 3D quantum gravity (Boulatov 1992). If we can crystallise these operations into a general algebraic process of quantisation it may be possible to construct a most complete algebraic structure by applying it multiple times starting from some arbitrary elementary algebra.

Putting it another way, we could say that the universe is described by the unique algebra which is the same as its own quantisation.

$$X_q \cong X$$